# Using dust particle vertical chains for wakefield investigation

Jie Kong, Truell W. Hyde, Jorge Carmona-Reyes, *Member, IEEE*

*Abstract*—**Dust particles often form vertical chains due to the influence of the wakefield produced by the streaming ions in the plasma sheath produced within an experimental dusty plasma. These particle chains are proving to be a unique diagnostic tool for investigating the physics behind the basic properties of the ion wakefield in the plasma sheath. In this paper, an experimental method is presented for investigating the wakefield employing a new oscillation technique designed to perturb a two-particle vertical chain. Damped "free fall" oscillations are generated via an external DC bias on the lower electrode. A fast Fourier transformation is used to produce a spectrum of the oscillation data collected and a center of mass oscillation is identified.**

## I. Introduction

Vertical chains of dust particles are often seen artifacts in laboratory dusty plasma experiments with the formation of two- or three-particle chains (or longer) under typical dusty plasma experimental settings (Fig. 1). In most cases, the lower particles within the chain appear to be controlled by the particles above them with the formation of such chains currently thought to be created by the ion wakefield. Ion wakefields are generated by the ion focusing effect created by ions streaming from the plasma to the lower electrode and passing through the floating dust particles [1-4]. Dust chain formation indicates that in addition to the gravitational, Coulombic or Yukawa forces acting on the particles, the ion wakefield force also plays an important role in establishing overall equilibrium within dusty plasmas. Therefore, a careful examination of the dynamic behavior of particle dust chains should lead to a better understanding of the attractive/repulsive

Manuscript received September 27, 2007. This work was supported in part by the National Science Foundation and the Department of Education.

Jay Kong is with the Center for Astrophysics, Space Physics & Engineering Research, Baylor University, Waco, TX. 76798 USA (e-mail: J_Kong@ baylor.edu).

Truell W. Hyde, is with the Center for Astrophysics, Space Physics & Engineering Research, Baylor University, Waco, TX. 76798 USA (phone: 254-710-3763, fax: 254-710-7309, e-mail: Truell_Hyde@baylor.edu).

Jorge Carmona Reyes is with the Center for Astrophysics, Space Physics & Engineering Research, Baylor University, Waco, TX. 76798 USA (e-mail: Jorge_Carmona@baylor.edu).

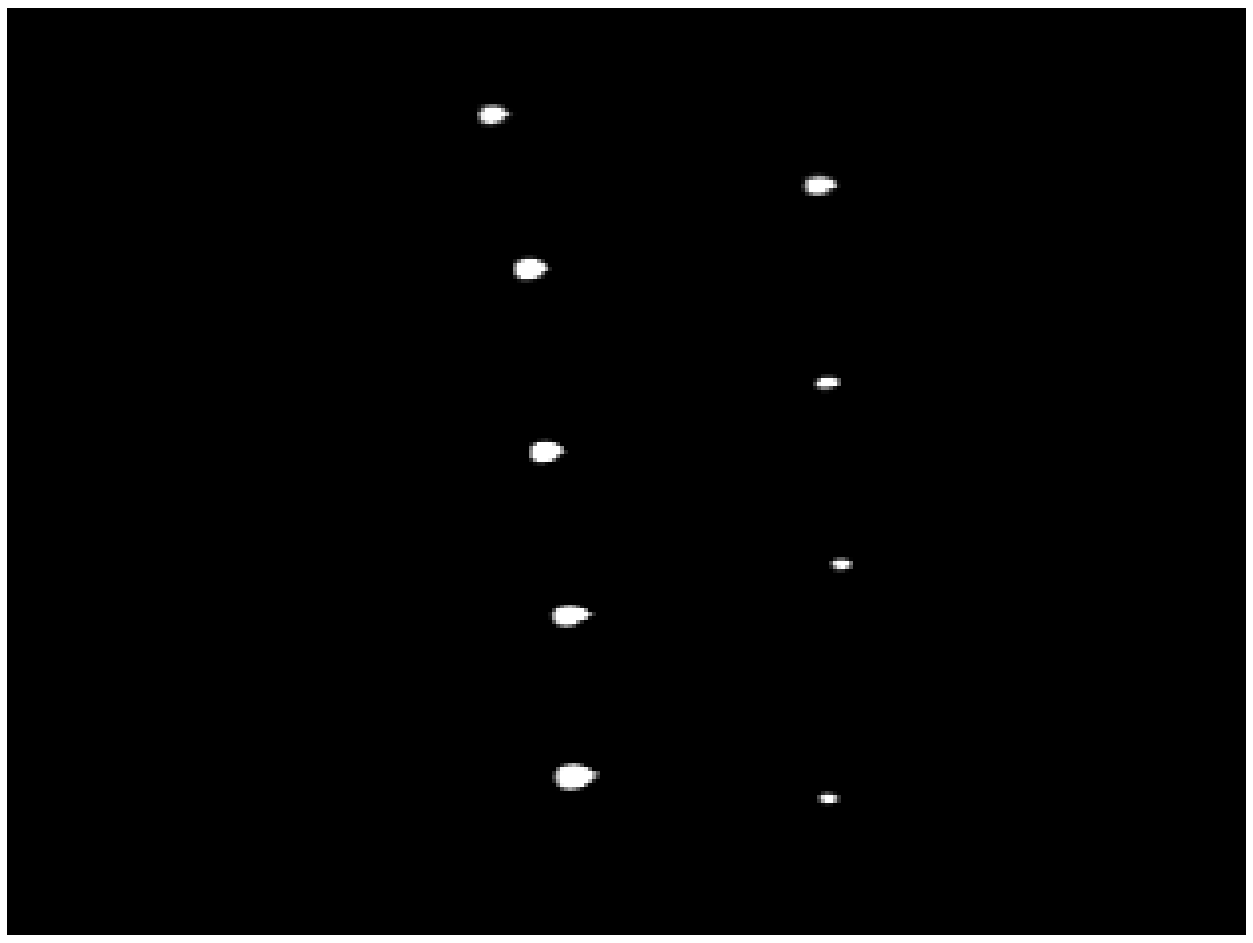

Fig. 1 Particle chains formed in a plasma sheath.

forces acting upon particles in dusty plasmas, particularly those in the vertical direction.

The shape and magnitude of the ion wakefield (and thus the behavior of any dust particles within) is directly related to the Mach number. The Mach number is defined as the ratio of the relative speed of the particle to the ion acoustic speed, $c_s = \sqrt{K_B T_e / m_i}$ , where $T_e$ is the electron temperature, $m_i$ is the mass of the ion and $K_B$ is the Boltzmann constant. Theoretical simulations [1] have shown that such ion wakefields form a potential well 'behind' the particle creating an attractive interparticle force. Recent experimental data show [5] that the strength of this attractive force is dependent upon the vertical separation distance between the particles, where this separation distance is also a function of the difference in particle size.

The above suggests that a two-particle chain provides an ideal mechanism for studying the ion wakefield. Experimentally, two-particle chains are relatively easy to produce under various laboratory settings and theoretically they have exact solutions. Therefore in this paper, a two-particle chain is employed to investigate the ion wakefield effect. The necessary data is collected by creating controlled oscillations within a two-particle chain, measuring the resulting relative motion between the particles and then employing a Fast Fourier transformation to provide a spectrum of the oscillations data.

## II. Two Particle Chain System

The governing equations of motion for a vertically aligned two-particle chain can be described by,

$$\ddot{y}_1 + \beta_1 \dot{y}_1 + \omega_{01}^2 y_1 = f_1(y_1 - y_2)$$
$$\ddot{y}_2 + \beta_2 \dot{y}_2 + \omega_{02}^2 y_2 = f_2(y_1 - y_2) \quad (1)$$

where for convenience, the numbers 1 and 2 are used to label the top and bottom particles, respectively. Thus, $y_{1,2}$ are their vertical positions, $\omega_{01,2}$ their resonance frequencies, $\beta_{1,2}$ the parameter related to their neutral drag coefficient, and $f_{1,2}$ describes the total interaction force between them. In this case, $f_{1,2}$ is assumed to include all forces (Yukawa, ion wakefield, etc.) which are a function of the relative distance between the particles.

To solve equation (1), it is advantageous to employ both the center of mass and relative frames of reference,

$$Y_{CM} = \frac{m_1 y_1 + m_2 y_2}{m_1 + m_2} = \frac{y_1 + \gamma y_2}{1+\gamma}$$
$$Y_R = y_1 - y_2 \quad (2)$$

where $\gamma = m_2/m_1$. Substitution of (2) into (1) yields,

$$\left(\ddot{Y}_{CM} + \beta_1 \dot{Y}_{CM} + \omega_{01}^2 Y_{CM}\right) + \left(\frac{\gamma}{1+\gamma}\right)\left(\ddot{Y}_R + \beta_1 \dot{Y}_R + \omega_{01}^2 Y_R\right) = f_1(Y_R) \quad (3a)$$

$$\left(\ddot{Y}_{CM} + \beta_2 \dot{Y}_{CM} + \omega_{02}^2 Y_{CM}\right) - \left(\frac{1}{1+\gamma}\right)\left(\ddot{Y}_R + \beta_2 \dot{Y}_R + \omega_{02}^2 Y_R\right) = f_2(Y_R) \quad (3b)$$

Equation (3) can be rewritten as,

$$\ddot{Y}_{CM} + \beta_1 \dot{Y}_{CM} + \omega_{01}^2 Y_{CM} = 0$$
$$\ddot{Y}_{CM} + \beta_2 \dot{Y}_{CM} + \omega_{02}^2 Y_{CM} = 0 \quad (4)$$

and

$$\left(\frac{\gamma}{1+\gamma}\right)\left(\ddot{Y}_R + \beta_1 \dot{Y}_R + \omega_{01}^2 Y_R\right) = f_1(Y_R)$$
$$-\left(\frac{1}{1+\gamma}\right)\left(\ddot{Y}_R + \beta_2 \dot{Y}_R + \omega_{02}^2 Y_R\right) = f_2(Y_R) \quad (5a)$$

Combining the above yields,

$$\ddot{Y}_R + \beta_0 \dot{Y}_R + \Omega_0^2 Y_R = f_1(Y_R) - f_2(Y_R) \quad (5b)$$

$$\Omega_0^2 = \frac{\gamma\omega_{01}^2 + \omega_{02}^2}{1+\gamma} \quad (5c)$$

where $\beta_0 = \frac{\gamma\beta_1 + \beta_2}{1+\gamma}$ and $\Omega_0$ and $\beta_0$ are the center of mass resonance frequency and center of mass neutral drag coefficient respectively.

As can be seen above, the mass difference between two particles in a chain can be determined using Equation (5b) once the individual particle's resonance frequency $\omega_{01,2}$ and center of mass resonance frequency $\Omega_0$ in the relative reference frame has been measured. Since the right hand side of equation (5b), $f_1 - f_2$, includes the ion wakefield force acting on particle 2, it can also be used to examine the wakefield effect in greater detail as will be shown below.

## III. Wakefield Potential Well

The wakefield potential downstream from a stationary dust particle can be analytically described as [1]:

$$\phi_w(r=0, y) \approx \frac{Q_D}{(4\pi\varepsilon_0 y)} \frac{2\cos(y/L_s)}{1 - M^{-2}} \quad (6)$$

where in the above, $L_s = \lambda_D \sqrt{M^2 - 1}$ is the effective length, $Q_D$ is the dust charge, M is the Mach number, r is the cylindrical coordinate, and the origin along the y direction is assumed to be at the current position of the dust particle.

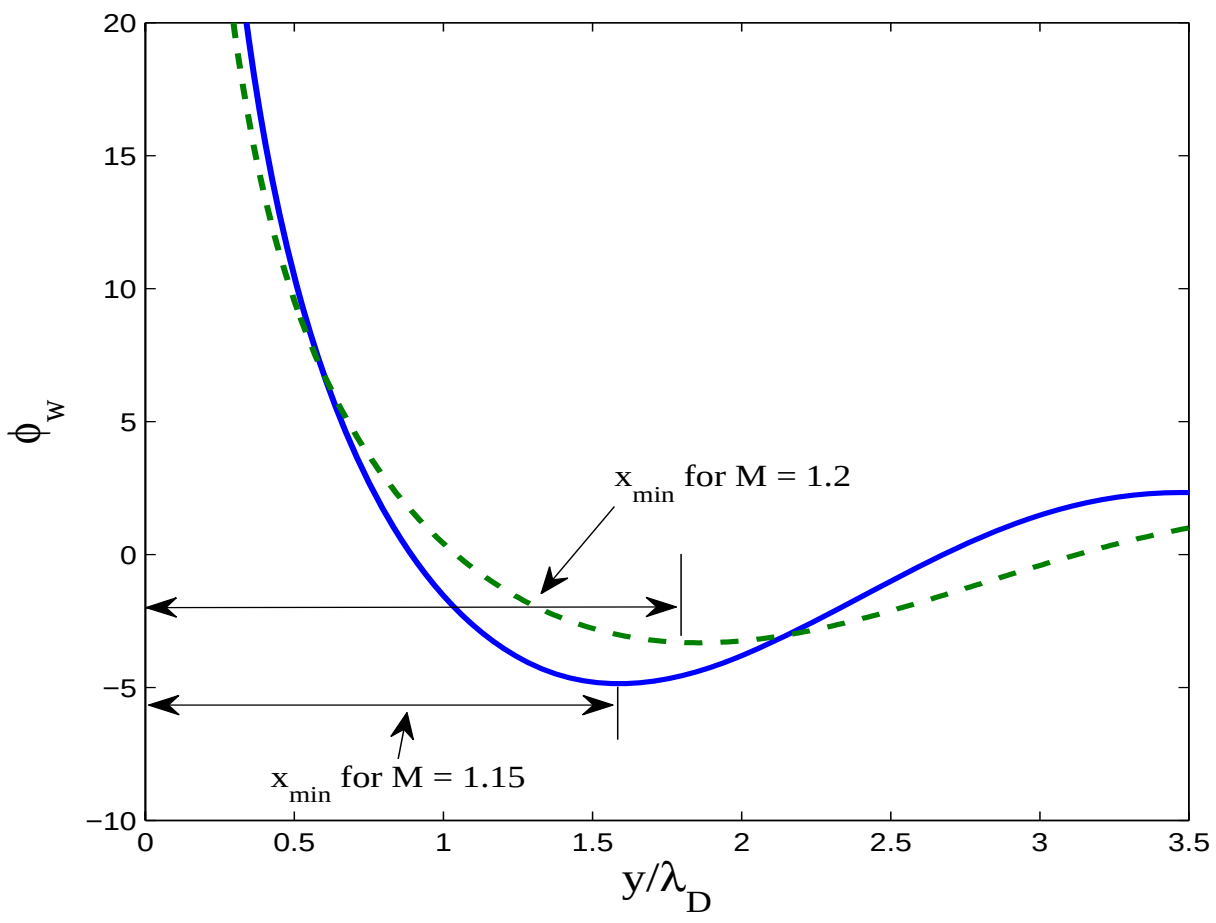

Fig. 2. The position of the wakefield potential well is a function of the Mach number. The smaller the Mach number, the closer the potential minimum is to the equilibrium position of the top particle. The separation between the top particle and the wakefield minimum is defined as $\Delta y = x_{min} \cdot \lambda_D$. (In the above, the top particle is assumed to be located at y = 0.)

The total interaction potential acting on the dust particle in addition to the sheath potential is then the sum of the wakefield and Yukawa potentials, $\phi = \phi_w + \phi_Y$. As shown in Fig. 2, $\phi_w$ has a primary minimum at $y/\lambda_D = x_{min}$, where this minimum is a function of M, $x_{min} = x_{min}(M)$.

Thus, any small displacement will create oscillations around this minimum where the resulting small amplitude vibrations within the potential well, $\phi_w$, can be expanded as a Taylor series around the point $y/\lambda_D = x_{min}$. Such expansion yields,

$$\phi(r=0,y)= A_0 + A_1\left(\frac{y}{\lambda_D} - x_{min}\right) + A_2\left(\frac{y}{\lambda_D} - x_{min}\right)^2 + \ldots \quad (7)$$

(Since small amplitude oscillations are assumed, terms of $O\left((y/\lambda_D - x_{min})^3\right)$ and above can be ignored.)

The resulting electrical field $E = -\frac{\partial\phi}{\partial y}$ is given by,

$$E(r=0,y) = -\left(2A_2\frac{y}{\lambda_D^2}\right) + C_0 \quad (8)$$

where $C_0$ is a constant. Therefore, the interaction force between dust particles in the chain is given by

$$F = Q_2 E = -Q_2\frac{2A_2}{\lambda_D^2}y + C_1 \quad (9)$$

where $C_1$ is a new constant. In the case of a second particle of mass $m_2$ and charge $Q_2$ located in this field, a perturbation from its equilibrium position will cause small oscillations with a resonance frequency of

$$\omega_w^2 = \frac{2Q_2A_2}{m_2\lambda_D^2} = \frac{Q_2}{m_2}\frac{Q_1}{\varepsilon_0\pi\lambda_D^3}M_f \quad (10)$$

where $M_f$ is a function of M, $M_f = M_f(M)$. Therefore, if the value of $\omega_w$ can be obtained experimentally, the Mach number of the system at the dust particle's position can be derived employing equation (10).

To evaluate the magnitude of $\omega_w$, equation (10) must be re-cast in the form,

$$\omega_w^2 = \omega_{02}^2\frac{Z_1M_f}{n_i\pi\lambda_D^3} \quad (11)$$

where $Q_1 = Z_1e$, $n_i$ is the ion density and $\omega_{02}$ is the lower particle's resonance frequency. Assuming normative dusty plasma experimental parameters ($Z_1 = 6\times10^3$, $K_BT_i = 0.03\text{eV}$ and $\lambda_D = 200\mu\text{m}$) yields $Z_1/n_i\pi\lambda_D^3 \approx 5$ and an oscillation frequency of $\omega_w \approx 2.2\omega_{02}\sqrt{M_f}$.

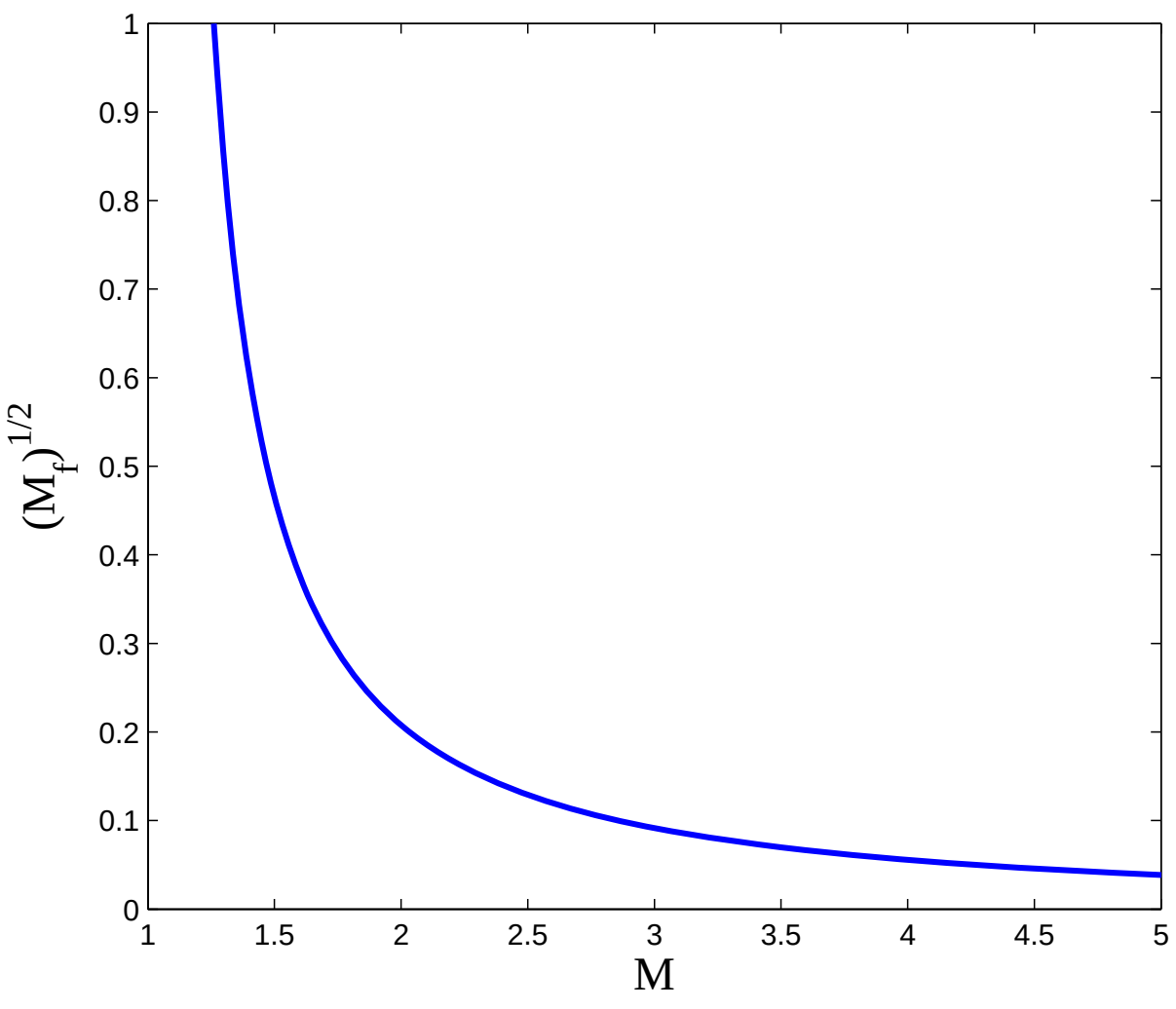

Fig. 3. $\sqrt{M_f}$ as a function of the Mach number. The monotonic nature of the relationship allows the calculation of $M_f$ from Equation (11) using experimentally determined data. Thus, the system's Mach number can be derived employing the above relationship.

Fig. 3 shows $\sqrt{M_f}$ as a function of M for all non-negative values of M. (Since $M_f$ is negative for $M < 1$, no oscillations occur in this region).

Rewriting equation (5b) by adding the derived wakefield oscillation term yields,

$$\ddot{Y}_R + \beta_0\dot{Y}_R + \Omega_{tot}^2 Y_R = f_{Yu}(Y_R) \quad (12a)$$

$$\Omega_{tot}^2 = \Omega_0^2 + \omega_w^2 \quad (12b)$$

where $f_{Yu}$ is the Yukawa force between the two particles. Equation (12) implies that the potential well created by the sheath and the potential well created by the wakefield overlap one another or are very close to doing so. Therefore, one necessary boundary condition for Equation (12) is that the lower particle must be positioned at or near the minimum of the wakefield potential. (It is important to reiterate that the balance between the Coulombic and gravitational forces will determine the equilibrium position in the vertical direction for the particle within the plasma sheath. The combined Yukawa and wakefield forces acting between the particles only produces small perturbations about this equilibrium position.) Fig. 4 shows two particles of different masses located within their individual sheath potential wells and separated by a distance $\Delta h$. Ignoring the interaction forces between the particles, the particle separation distance $\Delta h$ can be calculated

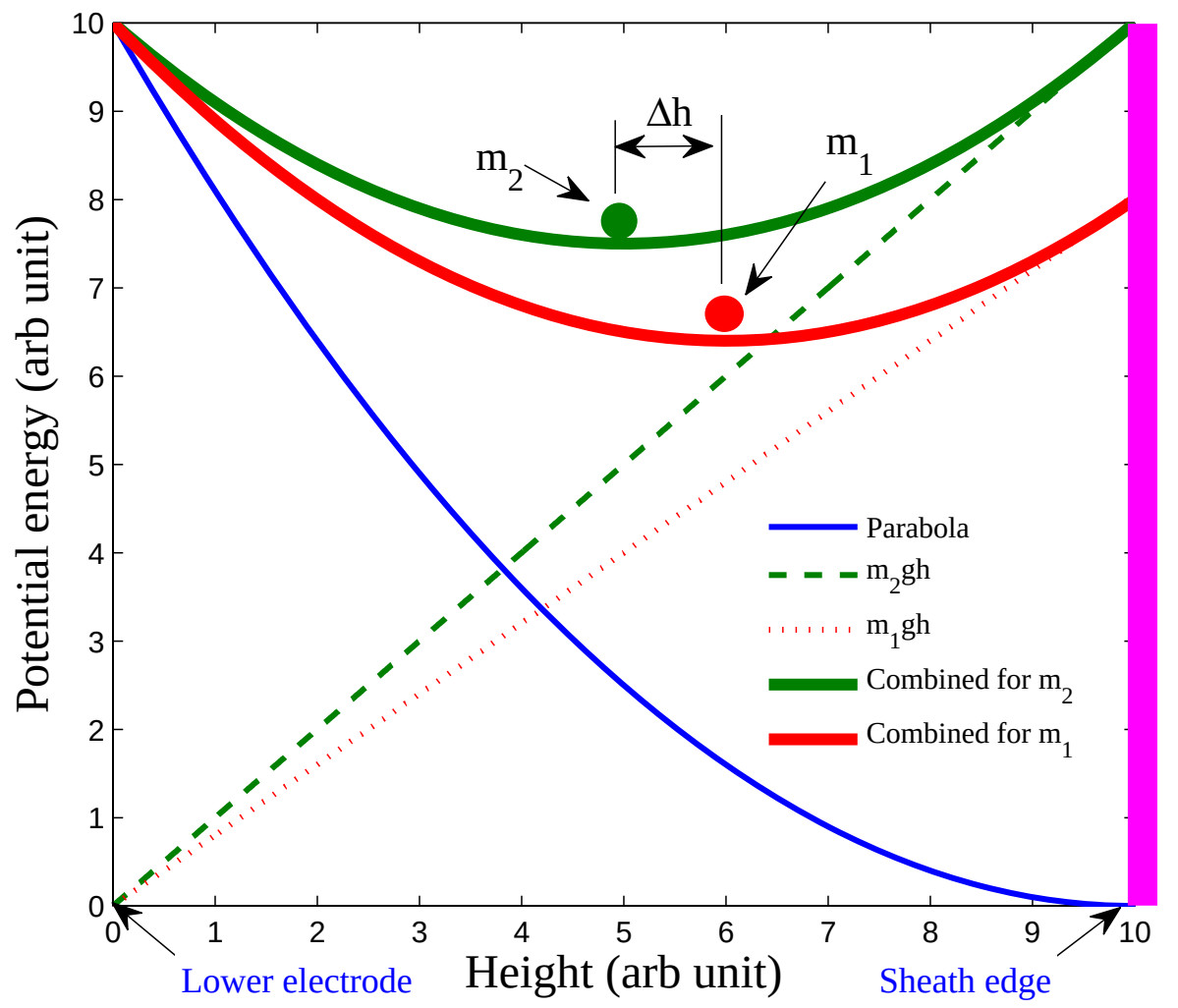

Fig. 4. Dust particles residing in individual 'wells' generated by the parabolic potential sheath and the gravitational potential. Δh is calculated using Equation (13) and data provided by measurement of the individual resonance frequencies of the particles.

employing [8],

$$\Delta h = \frac{m_2 g \varepsilon_0}{Q_2 n_i e} - \frac{m_1 g \varepsilon_0}{Q_1 n_i e} = \left( \frac{1}{\omega_{02}^2} - \frac{1}{\omega_{01}^2} \right) \cdot g \tag{13}$$

The wakefield potential minimum as determined from Equation (7), is located at a distance $\Delta y = x_{min} \cdot \lambda_D$ below the top particle, where $x_{min}$ is a function of the Mach number. (For example, $x_{min}$ = 1.6 at M = 1.15 for the previously given experimental values.) Therefore, the wakefield oscillation frequency $\omega_w$ only exists as defined by Equation (12) when $\Delta y \approx \Delta h$. When the minimum in the wakefield potential is located a distance away from the bottom particle's natural equilibrium position, i.e., $\Delta y >> \Delta h$, the effect of the wakefield in the vertical direction becomes more like that produced by a pure repulsive force. In this case, $\omega_w$ as defined in Equation (12) will not exist, i.e., $\omega_w = 0$.

## IV. Experimental Procedure and Results

The above technique was experimentally investigated employing the CASPER GEC rf reference cell [6]. In the CASPER GEC cell, a radio-frequency, capacitively coupled discharge is formed between two parallel-plate electrodes, 8 cm in diameter and separated by 3 cm, with the bottom electrode air-cooled. The lower electrode is powered by a radio-frequency signal generator, while the upper electrode is grounded as is the chamber. The signal generator is coupled to the electrode through an impedance matching network and a variable capacitor attenuator network. Positions for melamine formaldehyde particles of diameter $8.9 \pm 0.09$ µm were recorded using a CCD camera at 120 frames per second for an argon gas plasma held at 100 mTorr under a rf power of 5W at a frequency of 13.56 MHz. The CASPER plasma discharge apparatus is described in greater detail in [7].

Particles were perturbed by raising them a distance ΔH above their natural equilibrium position employing an external DC bias provided to the lower electrode, and then released by removing this external bias. This process created a series of attenuated oscillations around their natural equilibrium positions. A Fourier transformation was performed on the resulting oscillation data in order to obtain the frequency distribution spectrum. This procedure was then repeated at various values of ΔH, both positive and negative. Fig. 5 shows both the oscillation data and corresponding Fourier transformation spectra where an external bias of -10V from the natural bias was used with an RF power of 5W and an argon background pressure of 100 mTorr. As can be seen, the Fourier transformation spectrum for the lower particle (dotted line) shows a prominent peak at 15 Hz co-located with the theoretical peak. For the upper particle two peaks can be seen, one at 15 Hz and a second at 18 Hz. The larger of these is also the theoretical resonance frequency, while the peak at 15 Hz represents the interaction peak resulting from the resonance frequency of the bottom particle.

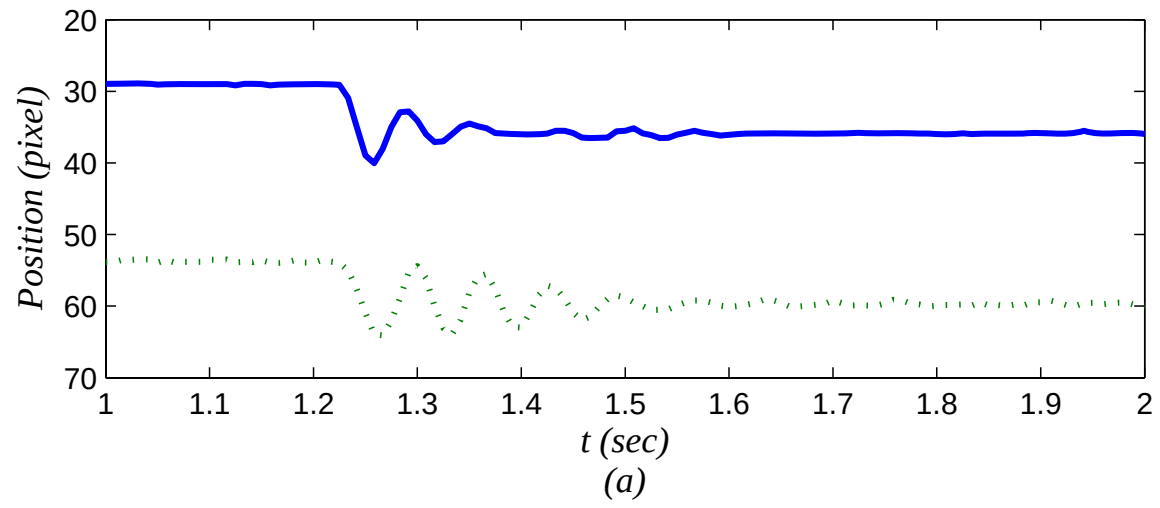

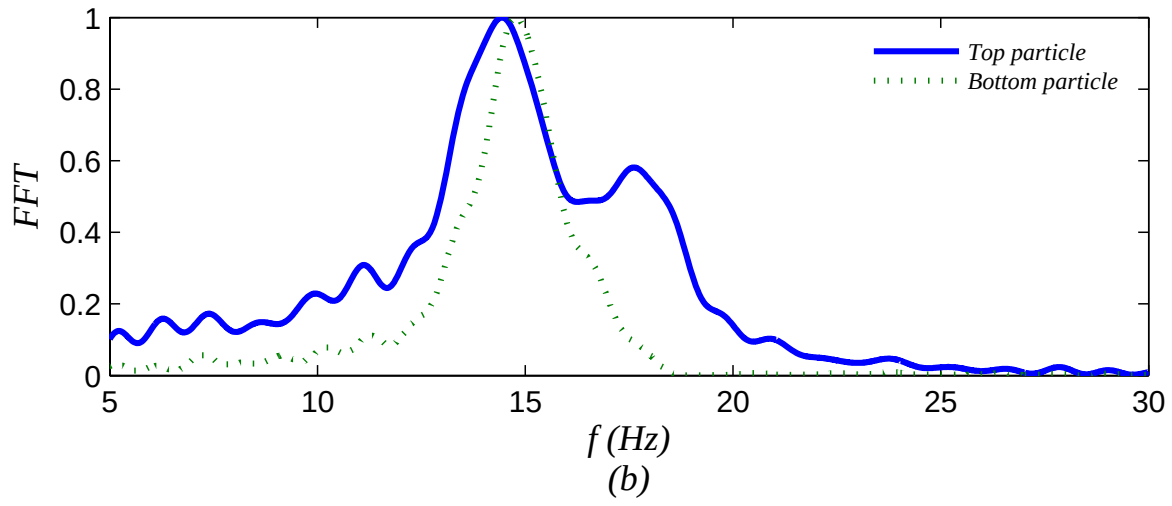

Fig. 5. (a) Particles undergoing damped oscillations after being released from a position ΔH above their natural equilibrium positions. (b) Fourier transformation spectra for the damped oscillation data shown in (a).

The above can be easily transformed from a frame showing only individual oscillations to one also showing relative oscillations, $Y_R = (y_1 - y_2) - a_0$, as denoted in Equation (5). Here a0 is the equilibrium separation between the two particles. Fig. 6 shows the relative oscillation data and its corresponding Fourier transformation spectrum. (The frequency step size is 0.01 Hz.) Three distinct peaks which can be attributed to $\omega_{02}$, $\Omega_{tot}$, and $\omega_{01}$ respectively can now be seen at 15.1 Hz, 16.4 Hz, and 17.7 Hz. (These are labeled as 1, 2, and 3 in the figure.) The theoretical fit, shown in Figure 6, assumes the amplitude of the damped oscillation as a

function

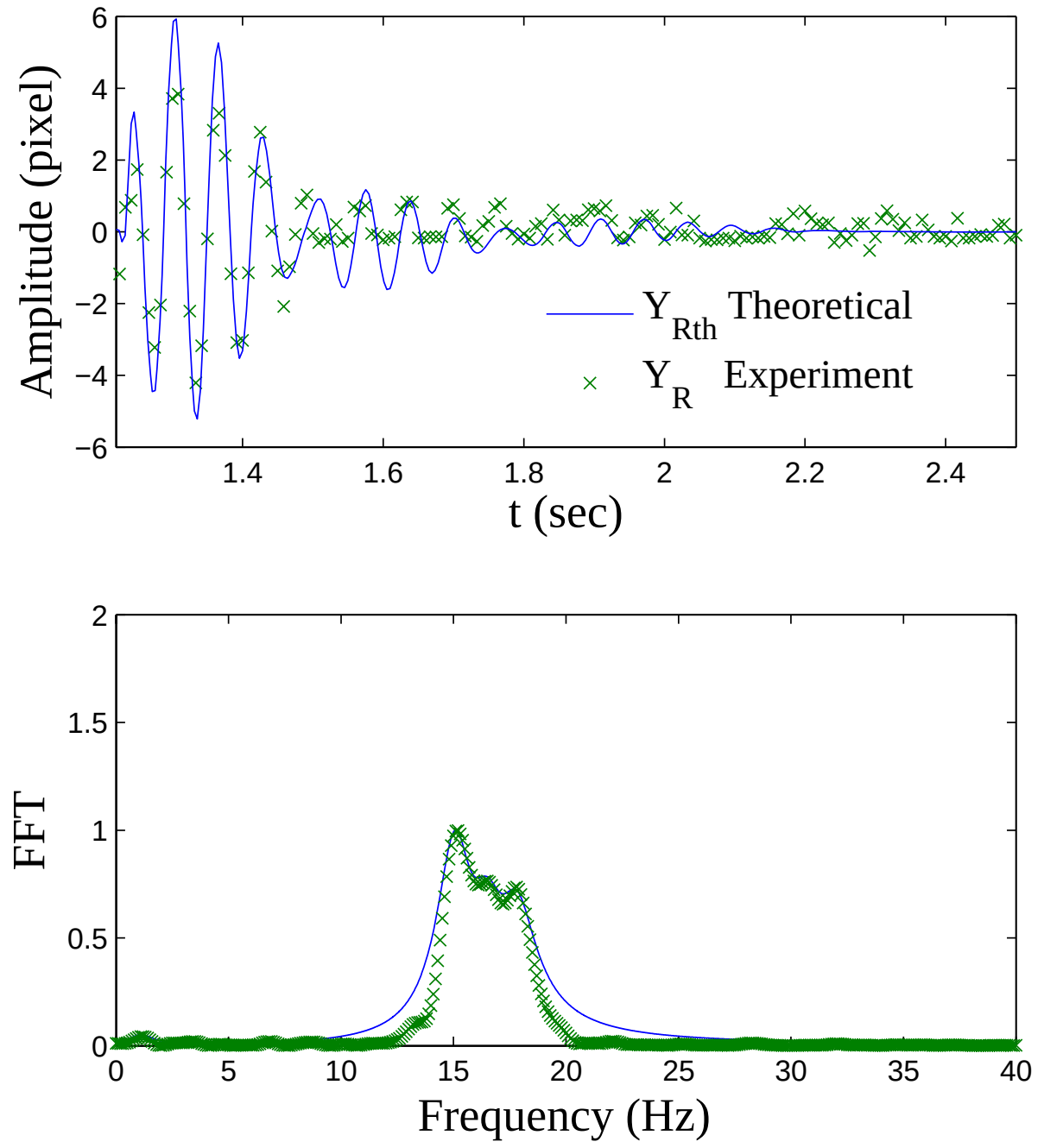

Fig. 6. Experimental data and theoretical fit for the relative separation between particles as derived from the data in Fig. 5. (a) Relative oscillation amplitudes as a function of time. $Y_R$ represents the experimental data while $Y_{Rth}$ shows a theoretical fit from data produced by Equation (15). (b) Fourier transformation of the data given in (a).

of time can be written as,

$$Y_{Rth} = \sum_i C(t) a_{0i} e^{-\beta_i t} \sin(\omega_i t) \qquad (15)$$

Here $C(t) = c_0 + c_1 t + c_2 t^2$, $c_0 = 0.15$, $c_1 = 1$, $c_2 = 5$ is an empirical modification factor derived from data fitting. Therefore, employing the experimentally measured values $\omega_{02}$, $\Omega_{tot}$, $\omega_{01}$, $\beta_{1,2,3} = 6\ s^{-1}$, and $a_{01} : a_{02} : a_{03} = 1 : 0.58 : 0.58$, the corresponding Fourier transformation spectrum can be determined after FFT($Y_{Rth}$).

## V. Discussion

For damped oscillation systems, the measured system frequency $\Omega_{tot}$ is

$$\Omega_{tot}^2 = \left(\Omega_0^2 + \omega_w^2\right) - \frac{\beta^2}{4} \qquad (16a)$$

yielding

$$\omega_w = \sqrt{\Omega_{tot}^2 - \frac{\gamma\omega_{01}^2 + \omega_{02}^2}{1+\gamma} + \frac{\beta^2}{4}} \qquad (16b)$$

where this equation has two unknowns, $\omega_w$ and $\gamma$. As pointed out earlier, for non-zero $\omega_w$ the particle separation $\Delta h$ and the wakefield potential minimum $\Delta y$ must be comparable, $\Delta h \sim \Delta y$. This restriction comes from the Taylor series expansion, where the bottom particle is assumed to be located at the wakefield potential minimum insuring small particle oscillation amplitudes. The experimentally measured average separation distance between the two particles examined in this research was 305 μm. Using $\omega_{01}$, $\omega_{02}$ and Equation (13), a calculated value for $\Delta h$ was determined to be $\Delta h = 290$ μm. In order to calculate the minimum for the wakefield potential, the Mach number for the system must first be determined. Previous experiments have shown that ions within the sheath can be easily accelerated to M = 3 but become saturated when M ~ 4 [9]. As shown in Fig. 2, in this case the calculated minima for the wakefield potential are at $3\lambda_D$, $6\lambda_D$ and $8\lambda_D$ for M = 2, 2.5, and 3 respectively, yielding $\Delta y$ values of 600, 1200 and 1600 μm respectively. The mass ratio $\gamma$ can be used to eliminate one unknown from Equation (16) but since the particle chain is formed naturally, $\gamma$ is not user definable and must be determined experimentally or given as a range using data provided by the manufacturer. Particles used in this experiment had measured diameters (as provided by the manufacturer) of $8.9 \pm 0.09$ μm, yielding a range for $\gamma$ of 1 − 1.06. According to Equation (11) and Fig. 3, $\omega_w$ increases as Mach number M decreases with the maximum $\omega_w$ as given by Equation (16b) at $\gamma = 1$.

Thus, using the experimental values previously determined for $\Omega_{tot}$, $\omega_{01}$, $\omega_{02}$, $\beta$ and assuming $\gamma = 1$, the calculated value for $\omega_w$ is approximately $0.5(2\pi)$ Hz, yielding a value of M > 3. This Mach number will create a wakefield potential minimum at about $\Delta y \sim 8\lambda_D$, or 1600 μm. Comparison of this value with the measured particle separation of 305 μm and the maximum relative oscillation amplitude of 250 μm, the condition $\Delta h \approx \Delta y$ is clearly not satisfied. Therefore, $\omega_w$ can be assumed to be 0 in Equation (16). When $\omega_w = 0$, a particle mass ratio $\gamma$ can be calculated; for this case, $\gamma = 1.01$ yielding a particle diameter ratio of 1.003. Assuming the top particle to have an average diameter of 8.90 μm, the bottom particle in this case must have a diameter of 8.93 μm. This value easily lies within the uncertainty range provided by the manufacturer ($\pm$ 0.09μm) and corroborates the overall method.

The Mach number is a system parameter, it does not depend on the particle size. Therefore, the condition of $\Delta h \sim \Delta y$ can be reached by adjusting the particle size difference based on Equation (13). For example, assuming the top particle is 8.9 μm, under the above experiment conditions the bottom particle should be around 13 μm to activate the wakefield oscillation frequency $\omega_w$.

## VI. Conclusions

It is generally accepted that the wakefield potential downstream from a dust particle in equilibrium within a dusty plasma can trap a second dust particle, forming a dust particle chain. This experiment shows that for a two-particle chain formed under these conditions, the separation distance between the particles is primarily determined by the sheath potential but that the center of mass oscillation frequency for the two-particle system is dependent upon either the sheath potential alone or the combined sheath and wakefield potential depending upon the conditions. When the lower particle is located near the wakefield potential minimum generated by the top particle (< $\pm$ one $\lambda_D$ length), the center of mass frequency for the two-particle system is determined from the combined sheath and wakefield potentials. However, when the lower particle is located some distance away from the wakefield minimum (> $\pm$ one $\lambda_D$ length) the center of mass oscillation frequency is determined by the sheath potential alone. As shown above, this allows either the mass difference between the two particles or the system's Mach number to be derived once the center of mass oscillation frequency has been determined.

**Jie Kong** received the B.S. degree in physics from Sichuan University, Chengdu, China, and the Ph.D. degree in surface analysis from Baylor University, Waco, TX. He is currently with Baylor University, where he is an Assistant Research Scientist in the Center for Astrophysics, Space Physics and Engineering Research. His research interests include complex (dusty) plasma diagnostics, plasma sheath, waves and instabilities, and phase transitions in complex (dusty) plasmas.

**Truell W. Hyde** (M'01) received the B.S. degree in physics from Southern Nazarene University, Bethany, OK, and the Ph.D. degree in theoretical physics from Baylor University, Waco, TX. He is currently with Baylor University, where he is the Director of the Center for Astrophysics, Space Physics and Engineering Research, a Professor of physics, and the Vice Provost for Research in the university. His research interests include space physics, shock physics and waves, and nonlinear phenomena in complex (dusty) plasmas.

**Jorge Carmona-Reyes** received the B.S. degree in physics from Southern Nazarene University, Bethany, OK, and the M.S. degree in physics from Baylor University, Waco, TX. He is currently with Baylor University, where he is an Assistant Research Scientist in the Center for Astrophysics, Space Physics and Engineering Research.